\documentstyle[aps,prl,epsf,twocolumn]{revtex}
\begin{document}

\draft

\wideabs{
\title{Reentrant charge ordering caused by polaron formation}

\author{Qingshan Yuan$^1$ and Peter Thalmeier$^2$}

\address{$^1$Max-Planck-Institut f\"{u}r Physik komplexer Systeme, 
N\"{o}thnitzer Str.38, 01187 Dresden, Germany\\
and Pohl Institute of Solid State Physics, Tongji University, 
Shanghai 200092, P.R.China\\
$^2$Max-Planck-Institut f\"{u}r Chemische Physik fester Stoffe, 
N\"{o}thnitzer Str.38, 01187 Dresden, Germany}

\maketitle

\begin{abstract}
Based on a two-dimensional extended Hubbard model with 
electron-phonon interaction, we have studied the effect of polaron 
formation on the charge ordering (CO) transition.  It is found that for fully 
ferromagnetically ordered spins the CO state may go through a process of 
appearance, collapse and reappearance with decreasing temperature. 
This is entirely due to a temperature-dependent polaron bandwidth. On the
other hand, when a paramagnetic spin state is considered, only a simple 
reentrant behavior of the CO
transition is found, which is only partly due to polaron effect.
This model is proposed as an explanation of the observed reentrant behavior of 
the CO transition in the layered manganite LaSr$_2$Mn$_2$O$_7$.

\end{abstract}


}

Electron crystallization was first proposed by Wigner, who considered an
electron system in a uniform positive background. 
It was pointed out that at sufficiently low densities the
Coulomb energy of electrons dominates their kinetic energy and then the 
electrons form a lattice. Such a Wigner lattice is experimentally realized 
in semiconductor heterostructures \cite{Andrei}.

Wigner's scenario is also important for solids 
\cite{Fulde}. Electron or hole crystallization, i.e., charge ordering (CO) is
extensively observed in real materials, even at high 
densities. The Verwey transition in magnetite Fe$_3$O$_4$ is a typical example
of a real space ordering of Fe$^{2+}$/Fe$^{3+}$ species \cite{Verwey}. 
Furthermore hole ordering is found in rare-earth pnictides like Yb$_4$As$_3$ 
because of intrinsic small hybridization between Yb-4f and As-4p orbitals 
\cite{Ochiai,FST}. The compound $\alpha'$-NaV$_2$O$_5$ which was originally 
proposed as another inorganic spin-Peierls material is now known to have a CO of 
V$^{4+}$/V$^{5+}$ at 34K \cite{Ohama}. CO also appears in a variety
of colossal magnetoresistance manganites, e.g., in La$_{1-x}$Ca$_x$MnO$_3$ 
for $x\ge 0.5$ \cite{Chen}. In all the above examples, only {\em one} CO 
transition is observed.

Recently, a melting of the CO state on decreasing the temperature,
i.e., reentrant behavior, has been found in layered manganite 
LaSr$_2$Mn$_2$O$_7$ by electron and x-ray diffraction \cite{Kimura}. 
It was proposed that a CO state exists characterized by superlattice reflections 
appearing  at $\sim 210$K, but collapsing again below $100$K. Subsequently, more 
precise measurements have found that surprisingly the superlattice intensities 
start growing
again below about $50$K meaning the reappearance of the CO state at lower 
temperatures \cite{Chatterji1}. That is to say, in the whole temperature region 
the CO state goes through a process of appearance, disappearance and 
reappearance. 

It is the purpose of this letter to propose a mechanism which can explain this 
intricate behaviour. It is known that the essence of CO transition is 
competition of the inter-site Coulomb energy and kinetic energy of electrons. 
Therefore a tempting idea is to invoke a temperature-dependent electron 
bandwidth such that the kinetic energy is small compared to the Coulomb energy 
at
some temperatures (favoring CO state), but dominates at other 
temperatures (favoring homogeneous state). Such a mechanism can be realized
by including electron-phonon interaction, i.e., considering
polaron formation since the effective hopping of a polaron is drastically 
affected by the number of thermally excited phonons. Indeed there is a lot of
evidence for polaron formation in manganite materials, for example in
La$_{1.2}$Sr$_{1.8}$Mn$_2$O$_7$ \cite{Romero}.
Very recently, small polaron ordering was directly observed in the CO phases in
La(Ca)MnO$_3$ and Pr(Ca)MnO$_3$ \cite{Li}. Theoretically the CO transition 
has already been studied within the polaron context \cite{Lee}, but more 
profound physics has been ignored. This will be addressed in the following. 
Rather than directly discussing material aspects we focus on a general
interesting problem: {\it how is the CO transition itself affected by 
a simultaneous polaron formation?}
The inherent evolution of the ordered polaron state with temperature is 
presented and rich reentrant behavior is found.

Our model is described by the following Hamiltonian in a two-dimensional 
square lattice:
\begin{eqnarray}
H & = & -t\sum_{i,\delta,\sigma}(c_{i,\sigma}^{\dagger}c_{i+\delta,\sigma}
+h.c.)+U\sum_{i}n_{i\uparrow}n_{i\downarrow} \nonumber\\
& & +V_1\sum_{i,\delta}n_i 
n_{i+\delta}+V_2\sum_{i,\eta}n_i n_{i+\eta} \nonumber\\
& & +f\sum_{i,\delta}u_{i,\delta}(n_{i+\delta}-n_i)+
\sum_{i,\delta}({P_{i,\delta}^2 \over 2m}
+{1\over 2} m\omega^2 u_{i,\delta}^2)\ ,
\end{eqnarray}
where $c_{i,\sigma}^{\dagger}$($c_{i,\sigma}$) denotes creation (annihilation)
operator for an electron at site $i$ with spin $\sigma$, $n_i$ is defined
as $n_i=n_{i\uparrow}+n_{i\downarrow}$ with $n_{i\sigma}=
c_{i,\sigma}^{\dagger}c_{i,\sigma}$, $t$ is 
nearest-neighbouring hopping parameter and $U,V_1,V_2$ represent on-site, 
n.n. and n.n.n. Coulomb repulsion, 
respectively. $\delta=\vec{x},\vec{y}$ are unit vectors along the $x$
and $y$ direction and $\eta=\vec{x}\pm \vec{y}$. 
The first four terms describe an extended Hubbard model. Here we consider the 
electron-phonon interaction in the following way. We assume that between 
arbitrary two n.n. sites there exists another kind of ion, 
e.g., oxygen, which vibrates along the bond
direction. Such a vibration will be coupled to the electron
density difference of two neighboring sites, as shown in the above
fifth term \cite{Song}. Then $m$ is the oxygen ion mass, $\omega$ is 
vibration frequency, $u_{i,\delta}$ represents the displacement of the
oxygen ion located between sites $i$ and $ i+\delta$. The vibrations have 
been assumed dispersionless for simplicity. Our electron-phonon coupling is
not of an usual on-site form which may be used to
model the coupling with the apical oxygen vibrations since
experiment on layered manganite La$_{1.2}$Sr$_{1.8}$Mn$_2$O$_7$ showed 
\cite{Romero} that polaronic distortions are constrained
within the perovskite layers \cite{Remark}. 
For further discussion, we rewrite the
electron-phonon interaction in the boson representation:
\begin{eqnarray}
H_{e-ph} & = & g\sum_{i,\delta}(b_{i,\delta}+b_{i,\delta}^{\dagger})
(n_{i+\delta}-n_i)+\omega\sum_{i,\delta}b_{i,\delta}^{\dagger}b_{i,\delta}
\ ,
\end{eqnarray}
where $g=f/\sqrt{2m\omega}$. We set $\hbar=k_B=1$ throughout and all energies
are in units of K.

The total number of sites is $N$ and for the electron band 
we assume quarter-filling. We will be interested in CO
with wavevector ($1/2,1/2$) and consider different occupancies on the two 
sublattices A and B of a bipartite lattice. First we treat the 
electron-phonon interaction with a combined Lang-Firsov (LF) transformation 
$U_1=\exp [-\sum_{i,\delta}g/\omega (b_{i,\delta}-b_{i,\delta}^{\dagger})
(n_{i+\delta}-n_i)]$ \cite{Lang}, and squeezing transformation 
$U_2=\exp [\gamma \sum_{i,\delta}(b_{i,\delta}b_{i,\delta}-
b_{i,\delta}^{\dagger}b_{i,\delta}^{\dagger})]$ with $\gamma$ ($>0$) 
a variational parameter \cite{Zheng}.
% i.e., $\tilde{H}=U_2 U_1 H U_1^{\dagger}U_2^{\dagger}$.
Then averaging on equilibrium phonon states \cite{Alexandrov}, we obtain
the following effective electronic Hamiltonian:
\begin{eqnarray}
H_{eff} 
& = & \omega (\tau +1/\tau) N/2 -\tilde{t}\sum_{i,\delta,\sigma}
(c_{i,\sigma}^{\dagger}c_{i+\delta,\sigma}+h.c.) \nonumber\\
& & +U\sum_{i}n_{i\uparrow}n_{i\downarrow}+V_1\sum_{i,\delta}n_i 
n_{i+\delta} \nonumber\\
& & +V_2\sum_{i,\eta}n_i n_{i+\eta}-g^2/\omega \sum_{i,\delta}(n_i-n_{i+\delta})^2 \ ,\label{Heff}
\end{eqnarray} 
where 
$\tau=\exp(-4\gamma)$ and $\tilde{t}=t\exp [-5\alpha \tau \coth (\omega /2T)]$ 
with definition of
dimensionless coupling constant $\alpha =g^2/\omega ^2$. It is obvious that 
electron-phonon interaction leads to two effects. One is that the 
bare hopping $t$ is reduced to $\tilde{t}$. The narrowing effect which is 
largely overestimated by the LF method alone in adiabatic limit 
($\omega /t <<1$) \cite{Mello} is weakened by the squeezing effect. 
Our emphasis here is the fact that $\tilde{t}$ itself 
increases with decreasing temperature because the phonon occupation decreases 
until only the contribution coming from zero-point
vibrations is left. Such a temperature-dependent renormalized hopping is very
important for reentrance of CO transition, as will be seen later. 
Another effect is phonon-mediated interaction 
between electrons which favors CO state, as described by the last term above. 
In our above treatment, 
the equilibrium positions of vibrating oxygens have been displaced, 
e.g., $\langle u_{i,\delta}\rangle\propto \langle n_i-n_{i+\delta}\rangle $,
which means the oxygens will be repelled to unoccupied sites once CO forms. 
In the following we concentrate on the CO transition for Hamiltonian
(\ref{Heff}). It is complicated if we simultaneously 
consider charge and spin degrees of freedom. We will therefore treat only 
cases with explicitly given spin configuration.

\begin{figure}
\epsfxsize=6.8cm
\epsfysize=4.2cm
\centerline{\epsffile{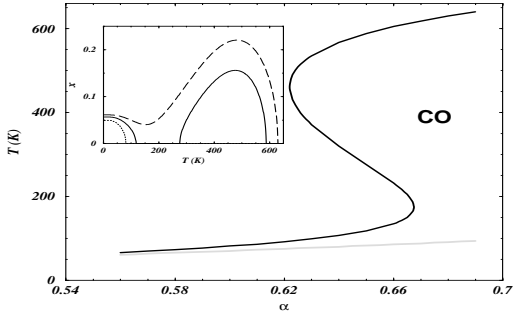}}
\caption{Phase diagram $T$ vs $\alpha$ for the case of fully FM spin order
with parameters $t=4000,V_{12}=400, \omega=200$.
The thin line is a hypothetical CO phase boundary for a constant 
$\tilde{t}=t\exp (-5\alpha \tau)$ (see text).
The inset shows the temperature-dependent order parameter $x$. The dotted 
line, solid line, dashed line correspond to $\alpha=0.60,\ 0.65,\ 0.68$, 
respectively. The multiple reentrant behavior is obvious for $\alpha=0.65$.}
\label{fig_phFM}
\end{figure}

First, we consider the fully ferromagnetical (FM) spin ordered case which is 
equivalent to the case of spinless electrons at half filling for which the 
Hamiltonian (\ref{Heff}), up to a constant, is simplified as 
\begin{eqnarray}
H_{eff} & = & \omega (\tau +1/\tau) N/2 -\tilde{t}\sum_{i,\delta}(c_{i}^
  {\dagger}c_{i+\delta}+h.c.)\nonumber\\
& &  +(V_1+2\alpha \omega)\sum_{i,\delta}n_in_{i+\delta} 
  +V_2\sum_{i,\eta}n_i n_{i+\eta} 
\end{eqnarray}
with $n_i=c_i^{\dagger}c_i$. In mean-field (MF) treatment 
and with the assumption $\langle n_i\rangle =1/2\pm x, i\in {\rm A,B}$
($x$ represents deviation from homogeneous electron distribution), 
the above Hamiltonian can be diagonalized with
energy bands $\pm \varepsilon _{\vec{k}}$. $\varepsilon _{\vec{k}}=
\sqrt{t_{\vec{k}}^2+(4Vx)^2}$ with $t_{\vec{k}}=4\tilde{t}\cos k_1\cos k_2\ 
(-\pi / 2<k_1,\ k_2\le \pi / 2)$ and $V=V_{12}+2\alpha \omega$
($V_{12}=V_1-V_2>0$) in the CO phase.
It is easy to find that in
this case the chemical potential $\mu =0$ for all temperatures, see also
Ref. \cite{Gebhard}. Note that it is only the difference $V_{12}$ of 
intersite Coulomb energies that enter the quasiparticle energy. 
Since $V_1,\ V_2$ are well screened and of 
the order $0.3$eV in manganites \cite{Mishra} the difference $V_{12}$ as 
estimated from 
neighbour distances is of the order $10^3$K. For this reason the CO transition
may occur in an observable temperature range. More precisely it is the 
difference of Madelung energies (with long range Coulomb interactions 
included) between the charge homogeneous and the CO phases that sets the 
scale of the transition temperature. Using the density of states for the 
spectrum $\varepsilon _{\vec{k}}$:
\begin{eqnarray}
D(\varepsilon) & = & {N\over 2\pi ^2}{\varepsilon \over \tilde{t}
\sqrt{\varepsilon ^2-(4Vx)^2}}K(\sqrt{1-z^2}) \ ,\nonumber
\end{eqnarray}
where $K$ is the complete elliptic function of the first kind and 
$z=\sqrt{\varepsilon ^2-(4Vx)^2}/4\tilde{t}$ we finally get the following
self-consistent equation for order parameter $x$ 
\begin{eqnarray}
1 & = & {2V\over \pi^2}\int_{0}^{1}{\tanh [2\sqrt{(\tilde{t}z)^2+(Vx)^2}/T]
\over  \sqrt{(\tilde{t}z)^2+(Vx)^2}}K(\sqrt{1-z^2})\ {\rm d}z \ . 
\end{eqnarray}
It can be easily proven that this equation always has a solution
with nonzero $x$ at zero temperature no matter what are the values of 
parameters $t,\alpha,V_{12}$. In the current half-filled spinless 
case this means that the system must take on CO state at zero temperature. We 
will refer to this fact later on.
For given initial values for $t,V_{12},\omega$ (for example, 
$t=4000,V_{12}=400,
\omega=200$), we give the solution of the above equation in the phase diagram
$T$ vs $\alpha$, shown in thick line of Fig. \ref{fig_phFM}. The parameter
$\tau$ (or $\gamma$) is determined by minimization of ground state energy of
the homogeneous state. Its minor variation with parameter $\alpha$ is
ignored since it does not change our qualitative results. An average value 
$\tau=0.12$ is adopted.
From the phase diagram it can be clearly seen that there is an intermediate
region for $\alpha$ where the system goes through a process of 
homogeneous state, CO
state, homogeneous state and CO state with decreasing temperature. The
corresponding variation of the order parameter $x$ with temperature is shown 
in the inset of Fig. \ref{fig_phFM} by a solid line.  
The physical explanation is as follows. At high temperature, the hopping
is drastically reduced so that CO state is favored to appear at a
relatively high temperature (It is usually quantitatively  
overestimated due to our simple treatment). 
On decreasing temperature the effective hopping
increases. And most importantly, it increases so fast that 
its corresponding CO 
transition temperature decreases even faster than the temperature
itself. Therefore the CO state cannot be reached any more and the system 
returns to the homogeneous state. However, at even lower temperature, 
the system is sure to recover to CO state since the ground state of the system
is charge ordered, as refered above. Actually such a
conclusion is strongly favored by other methods beyond MF theory 
\cite{Gebhard,Shankar}. Physically one has a half-filled band of spinless
electrons with perfect nesting which has a CO instability for arbitrary
small inter-site Coulomb interaction at zero temperature.
In this sense it is inevitable that the 
system finally enters the CO state. 
We want to emphasize, the multiple reentrant behavior found here is 
entirely due to a temperature-dependent polaron bandwidth. For comparison, 
a hypothetical
phase diagram without the temperature-dependent factor in hopping $\tilde{t}$,
i.e., $\tilde{t}=t\exp (-5\alpha \tau)$ 
%(an equivalent circumstance to such a case will be given later)
is shown as thin line in Fig. \ref{fig_phFM}. Obviously no reentrance of CO 
happens now.

The phase diagram with its reentrant behavior is similar for 
different values of parameters. However it is appropriate to discuss the 
influence of the optical phonon frequency $\omega$.
It is obvious that the renormalized hopping $\tilde{t}$ is nearly 
unchanged with the temperature in the region about $T<\omega/2$ since 
the phonons are difficult to excite. This means
that the temperature-dependent effective hopping which we
emphasized here is mainly observed in the region $T>\omega/2$.
Therefore, roughly speaking, if we increase the frequency the reentrant 
region will as a whole move to higher temperatures.
It should be also clearly pointed out that our temperature dependence of 
the effective hopping $\tilde{t}$ is not quantitatively exact. However the 
precise form of $\tilde{t}(T)$ is not crucial for the above reentrance 
behavior as has been checked \cite{Comment} with a 
functional dependence of $\tilde{t}(T)$ less pronounced than the exponential
behavior given before.

\begin{figure}
\epsfxsize=6.8cm
\epsfysize=4.2cm
\centerline{\epsffile{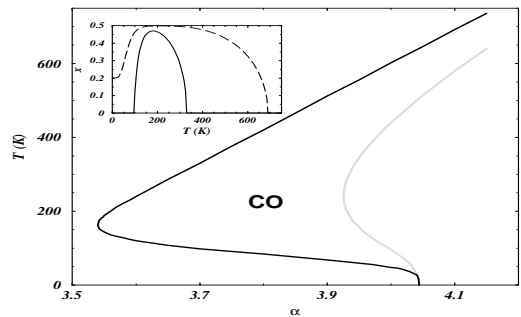}}
\caption{Same to Fig. \ref{fig_phFM} for the case of PM spin order
with parameters $t=4000, V_{0}=-2000, \omega=200$. In the inset the 
solid and dashed lines correspond to $\alpha=3.7,\ 4.1$, 
respectively. Only one reentrant transition is obvious for $\alpha=3.7$.}
\label{fig_phPM}
\end{figure}

Now, we discuss the case including spin and assume a paramagnetic (PM) spin 
state with $\langle n_{i\uparrow}\rangle =\langle n_{i\downarrow}\rangle 
=\langle n_i\rangle /2$. For moderate $U$
and with similar steps as before we can obtain self-consistent MF
equations for the order parameter $x$ and chemical potential $\mu$:
\begin{eqnarray}
1 & = & {4V \over \pi^2}\int_{0}^{1}{1 \over \sqrt{(\tilde{t}z)^2+(Vx)^2}}
F_{-}(z)K(\sqrt{1-z^2})\ {\rm d}z \ ,\label{PMx}\\
1 & = & {8 \over \pi^2}\int_{0}^{1}
F_{+}(z)K(\sqrt{1-z^2})\ {\rm d}z  \label{PMmu}
\end{eqnarray}
with functions $ F_{\pm}(z)=
 1/(1+\exp [-(4\sqrt{(\tilde{t}z)^2+(Vx)^2}+\mu)/T]) \pm
1/(1+\exp [(4\sqrt{(\tilde{t}z)^2+(Vx)^2}-\mu)/T])  $
and the new definition $V=V_0+3\alpha \omega$ ($V_0=V_{12}-U/8$). At zero
temperature it is easy to obtain the condition:
$V/\tilde{t}> {\rm const} \approx 1.2$, 
under which a solution with nonzero $x$ exists. Obviously 
$\alpha$ must exceeds a threshold to reach
CO state as long as initially $V_0<1.2t$. This is a major difference from the
fully FM spin ordered case. The numerical solution for Eqs. (\ref{PMx}) and
(\ref{PMmu}) is summarized in the phase diagram of Fig. \ref{fig_phPM} with 
a thick line (we have set the same value for $\tau$ as before). 
Due to different zero temperature properties now only
possibility of {\em one} reentrance is found as shown by the solid line in the 
inset. For comparision, 
we also give the hypothetical phase diagram with
$\tilde{t}=t\exp (-5\alpha \tau)$ by
a thin line in Fig. \ref{fig_phPM}. Reentrant
behavior is also found. Actually, our hypothetical case is qualitatively
equivalent to considering a $T$ vs $V_{12}$ (with fixed $U$) phase diagram 
when electron-phonon
interaction is not included. Even without electron-phonon interaction,
reentrant behavior is also possible in some $V_{12}$ region in the PM spin 
case, see also Ref. \cite{Pietig}. However, the quantitative reentrant 
behavior is much different with and without polaron effect. 
%The latter case was also considered in Ref. \cite{Pietig}. 

It is natural to think about a crossover between the two extreme spin 
cases, PM and complete FM, i.e. to consider the case of general incomplete 
FM spin order. Since the fully FM spin order can be considered as a 
limit of on-site repulsion $U\rightarrow \infty$,
such a crossover has to be treated beyond MF which will be investigated in the 
future. We predict that even in the intermediate case the multiple reentrant 
behavior is controlled by the polaron effect.
The extended Hubbard model itself is expected to show at most one reentrant 
transition. This conjecture is obtained from an interpolation between Fig.1
and 2, see the thick and thin phase boundaries respectively.
 
Finally we give a possible explanation of the experimental
observations in 
manganite LaSr$_2$Mn$_2$O$_7$ in the context of our theory. This material is 
composed of
MnO$_2$ bilayers and there is one $e_g$ electron every two manganese ions.
When double-exchange mechanism is considered, the initial hopping $t$ should
be substituted by $t\langle \cos (\theta /2)\rangle $ ($\theta$: angle between
two $t_{2g}$ spins on n.n. sites), which depends on local spin order. 
There is strong evidence \cite{Kubota} that in this material in-plane FM 
spin order has already been well developed within the MnO$_2$ layers at 
room temperature, before it exhibits true long range order. This is also
concluded from the observation of FM in-plane spin waves even at twice the
3D transition temperature \cite{Chatterji2}.
So a good assumption is considering that the spins (for both $t_{2g}$ 
and $e_g$ electrons) are perfectly aligned in the whole
temperature region as far as a single layer is concerned. For the same 
reason the dependence of $t$ on the temperature resulting from double-exchange 
mechanism is ignored. Then we can use our theory with fully FM 
spin order to explain the CO scenario of LaSr$_2$Mn$_2$O$_7$. 
The reentrant behavior 
seen in Fig. \ref{fig_phFM} is qualitatively the same as found
experimentally \cite{Chatterji1}. Moreover the experimental observation of 
reappearing superlattice intensities at lower
temperatures found in Ref.\cite{Chatterji1} is naturally explained within 
our theory. We point out, the multiple reentrant behavior 
is also possible from our theory even if the assumption about
perfect FM spin order is relaxed. Of course, full explanation on the 
experiments needs to consider other factors, especially an orbital degree of 
freedom \cite{Kubota}. However, we think that polaron formation plays an 
essential role for the observation of reentrant CO transitions 
in this material.

In conclusion, in a 2D extended Hubbard model we have found interesting 
reentrant behavior of CO transition when including the polaron effect. 
Due to a 
temperature-dependent polaron bandwidth the CO state may go through a process 
of appearance, collapse and reappearance on decreasing the temperature in case
of fully FM spin order. When a PM spin order is considered, once 
reentrance of CO
transition may happens, which is partly due to polaron
effect. The crossover between these two cases is complicated and still has 
to be investigated.

Q. Yuan would like to thank P. Fulde for his encouragement and
R. Pietig, H. Zheng, F. Gebhard for helpful discussions or communications. 
P. Thalmeier would like to thank 
T. Chatterji for cooperation and discussions.


\begin{references}
\bibitem{Remark} If the on-site Holstein model is used the renormalized
hopping $\tilde{t}$ in Hamiltonian (\ref{Heff}) has a different numerical 
constant in the exponent. However this does not change the qualitative 
results about the reentrant CO transition.
\bibitem{Comment} To prove this conjecture we used an empirical effective 
hopping $\tilde{t}=t/(1+(T/\omega)^n)$ with qualitatively correct 
behavior for low and high temperatures but a much weaker power law 
$T$-dependence than obtained here (cf. expression below Hamiltonian 
(\ref{Heff})). For an exponent $n>1.5$ we always observe the same 
reentrance transitions.
\end{references}
\end{document}